\documentclass[twoside]{article}
\usepackage{fleqn,espcrc2}
\usepackage{graphicx}
\usepackage{amssymb}

\newcommand{\AmS}{{\protect\the\textfont2
  A\kern-.1667em\lower.5ex\hbox{M}\kern-.125emS}}

\title{A frequentist analysis of solar neutrino data\thanks{
Talk presented by M.V. Garzelli at the
EuroConference on \emph{Frontiers in Particle Astrophysics and Cosmology},
San Feliu de Guixols, Spain, 30 Sep. -- 5 Oct. 2000;
DFTT 45/00.
}}

\author{M.V. Garzelli and C. Giunti\\[0.2cm]
INFN, Sez. di Torino, and Dip. di Fisica Teorica,
Univ. di Torino, I--10125 Torino, Italy}
       
\begin{document}

\begin{abstract}
We present a Monte Carlo analysis
in terms of neutrino oscillations
of the total rates measured in solar neutrino experiments
in the framework of frequentist statistics.
We show that the goodness of fit
and the confidence level of
the allowed regions in the space of the neutrino oscillation parameters
are significantly overestimated in the standard method.
We also present a calculation of exact allowed regions
with correct frequentist coverage.
We show that
the exact VO, LMA and LOW regions are much larger than the standard ones
and merge together
giving an allowed band at large mixing angles
for all $\Delta{m}^2 \gtrsim 10^{-10} \, \mathrm{eV}^2$.
\end{abstract}

\maketitle

\section{INTRODUCTION}

Solar neutrino experiments~\cite{sunexp}
have observed a flux of neutrinos
smaller than the one predicted
by the Standard Solar Model
(see, for example, Ref.\cite{Bahcall}).
Neutrino oscillations
(see, for example, Ref.\cite{BGG-review-98-brief})
is widely considered to be
the simplest and most attractive explanation
of this anomaly.
Assuming the simplest case of two-neutrino oscillations,
the statistical analysis of solar neutrino data
yields allowed regions for the oscillation parameters
$\Delta m^2 \equiv m_2^2 - m_1^2$,
where $m_1$ and $m_2$ are the two neutrino masses,
and
$\tan^2\theta$,
where $\theta$ is the neutrino mixing angle.

The allowed regions in the
$\tan^2\theta$--$\Delta m^2$ plane
are usually
determined through a least-square fit
(see, for example, Refs.\cite{dard,faci}),
in which the parameters are estimated through
the minimum of the function
\begin{eqnarray}
X^2
&=&
\sum_{j_1,j_2=1}^{N_{\mathrm{exp}}}
\left( R^{\mathrm{(thr)}}_{j_1} - R^{\mathrm{(exp)}}_{j_1} \right)
(V^{-1})_{j_1j_2}
\nonumber
\\
&&
\hspace{1cm}
\times
\left( R^{\mathrm{(thr)}}_{j_2} - R^{\mathrm{(exp)}}_{j_2} \right)
\,,
\label{X2}
\end{eqnarray}
where
$N_{\mathrm{exp}}=3$
is the number of experimental data points,
$R^{\mathrm{(thr)}}_{j}$
and
$R^{\mathrm{(exp)}}_{j}$
are the theoretical and experimental rates,
respectively,
and
$V$ is the covariance matrix that accounts for experimental and theoretical 
uncertainties~\cite{stan,dard,noi-cs}.
The theoretical rates $R^{\mathrm{(thr)}}_{j}$
and the covariance matrix $V$
depend on the parameters $\Delta m^2$, $\tan^2\theta$.

Usually,
in the application of the least-squares method it is assumed
that $X^2$ is distributed 
as a $\chi^2$ with $N_{\mathrm{exp}}=3$ degrees of freedom,
$X^2_{\mathrm{min}}$
is distributed as a $\chi^2$
with $N_{\mathrm{exp}}-N_{\mathrm{par}}=1$ degrees of freedom,
and
$X^2-X^2_{\mathrm{min}}$
has a $\chi^2$ distribution with $N_{\mathrm{par}}=2$ degrees of freedom
(see, for example, Ref.\cite{Eadie-71,Numerical-Recipes}).
This would be correct if:
1)
the theoretical rates depended linearly on the parameters;
2)
the errors of the differences
between the theoretical and experimental rates were
multinormally di\-stri\-bu\-ted with a constant covariance matrix.
Actually, these requirements are not satisfied.
In particular,
it is well-known that
the theoretical rates have a complicate 
dependence on the parameters.
For example,
in the simplest case of
oscillations in vacuum the electron neutrino survival probability depends
on $\Delta m^2$ through a sinusoidal function:
\begin{equation}
P_{\nu_e\to\nu_e}
=
1 - \sin^2{2\theta}\sin^2{\left(\Delta m^2 L/4E\right)}
\,.
\end{equation}
Moreover,
the covariance matrix $V$ is not constant,
but depends on
$\Delta m^2$ and $\tan^2\theta$
(see Refs.\cite{stan,dard,noi-cs}),
and
the errors of the differences
between the theoretical and experimental rates are not
multinormally distributed
(this is due to the fact that
each theoretical rate is given by the product
of the neutrino flux times the experimental cross section;
even if
the errors of the neutrino flux and the errors of 
the experimental cross section
are normally distributed,
their product is not).

Since $X^2$ is {\it not} a $\chi^2$,
in order to perform a reliable statistical analysis of the data
it is necessary
to calculate with a Monte Carlo the distribution of its minimum,
$X^2_{\mathrm{min}}$,
which is the estimator of the parameters
$\Delta m^2$, $\tan^2\theta$.

In Section~\ref{GOF}
we present the result of our Monte Carlo
estimation of the goodness of fit and the confidence level (CL)
of the standard allowed regions.
In Section~\ref{EXACT}
we present a calculation of
exact allowed region with correct frequentist coverage.
A detailed explanation of our procedure and results
has been presented in Ref.~\cite{noi-mc}.
  
\begin{figure*}[t!]
\begin{center}
\rotatebox{90}{\includegraphics[bb=100 80 540 741,width=7cm]{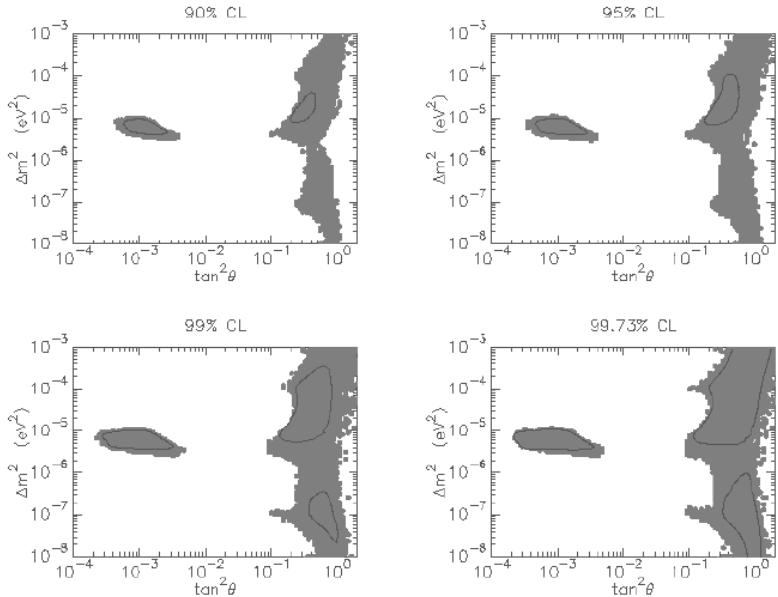}}
\end{center}
\caption{Allowed $90\%$, $95\%$, $99\%$, $99.73\%$ CL regions in the
MSW part of the
$\tan^2{\theta}$--$\Delta m^2$ plane.
The gray areas are the allowed regions with exact frequentist co\-ve\-ra\-ge in
the MSW region; the areas enclosed by the solid lines are the standard SMA, LMA
and LOW allowed regions.}
\label{fig:mswqua}
\end{figure*}

\section{MONTE CARLO GOODNESS OF FIT AND CONFIDENCE LEVELS}
\label{GOF}

The exact distribution of
$X^2_{\mathrm{min}}$
is determined by the true value of the parameters
$\Delta m^2$, $\tan^2\theta$,
which are unknown.
Nevertheless,
it is possible to estimate the distribution
assuming surrogates for the true value of the parameters.
The most reasonable surrogates
are the best-fit values
$\widehat{\Delta m^2}$ and $\widehat{\tan^2{\theta}}$
(see, for example, Ref.\cite{Numerical-Recipes}).
Assuming $\widehat{\Delta m^2},\widehat{\tan^2{\theta}}$
as surrogates of the true values of the parameters,
we generate
$N_S$ synthetic data sets which simulate $N_S$ different independent
sets of experiments.
Using these sets we can estimate the goodness of fit and confidence levels
of the standard allowed regions.

{\it Goodness of fit} is the probability
to find
in a set of hypothetical repeated experiments
a $X^2_{\mathrm{min}}$
larger than the one actually observed.

From the rates analysis
(with the 1999 data summarized in Ref.\cite{faci}),
we find that,
constraining the parameter in an area around the global minimum
(which is the SMA region),
the standard
goodness of fit is reliable.
This is due to the fact that
in the neighborhood of the global minimum
the dependence of the theoretical rates from the parameters
is approximately linear
and the covariance matrix is almost constant.

On the other hand,
if the parameters are unconstrained
(allowing all the MSW region with
$
10^{-8} \, \mathrm{eV}^2
\lesssim
\Delta{m}^2
\lesssim
10^{-4} \, \mathrm{eV}^2
$
and the VO region with
$
10^{-11} \, \mathrm{eV}^2
\lesssim
\Delta{m}^2
\lesssim
10^{-8} \, \mathrm{eV}^2
$),
the goodness of fit is 40\%,
poorer than the 52\% obtained with the standard method.
The fit is worse!
We conclude that the
goodness of fit is {\it overestimated} by the standard procedure.

From the synthetic data sets we can also estimate
the confidence level of the standard
allowed regions.
We start from the definition of {\it confidence level} of
interval:
it is the fractional number of intervals
obtained in repeated experiments
which cover the true
values of parameters.

In the standard procedure the confidence intervals are 
determined by the condition
\begin{equation}
X^2
\leq
X^2_{\mathrm{min}}+\Delta \chi^2(\beta)
\,,
\end{equation}
where $\Delta \chi^2(\beta)$ is the value of $\chi^2$ such that the cumulative 
$\chi^2$ distribution for a number of degrees of freedom equal to the
number of parameters is equal to the confidence level $\beta$.
But this procedure would be correct only if $X^2$ were a $\chi^2$.

In our simulation,
for a fixed $\beta$,
we  calculate the standard allowed regions
for each synthetic data set.
Then we count how many of them cover
the surrogates of true values of parameters.
This fraction is our estimate $\beta_{\mathrm{MC}}$
of the true confidence level
of the standard allowed regions at $\beta$ CL.

We found that,
if the parameters are constrained around the global minimum
and there is only one standard $\beta$ CL allowed region,
its confidence level $\beta_{\mathrm{MC}}$
is approximately equal to $\beta$.
Instead,
if the parameters are not constrained around the global minimum
and
there are several standard $\beta$ CL allowed regions,
their confidence level $\beta_{\mathrm{MC}}$
is significantly smaller than $\beta$.
For example,
the Monte Carlo confidence level
of the standard 90\% CL allowed regions
is only 86\%.

This result was expected,
because
in the case of a real
$\chi^2$
there is only one mi\-ni\-mum that determines
an elliptic allowed region,
whereas $X^2$
has several local mi\-ni\-ma
(especially in the vacuum oscillation region).
Therefore,
in repeated experiments
there is a higher probability that the global minimum
falls far from the assumed surrogates
$\widehat{\Delta m^2}$,
$\widehat{\tan^2{\theta}}$
of the true values of parameters,
leading to a lower probability
that the allowed regions cover 
$\widehat{\Delta m^2}$,
$\widehat{\tan^2{\theta}}$.

The approach presented in this section
allows to calculate only estimations
of the goodness of fit and
the confidence level of the standard allowed regions,
because the calculation
depends on the assumed surrogates of the true values of the parameters.
In the next section
we present the results of a calculation of allowed regions
with exact confidence level,
which is independent from the unknown true value of the parameters.

\section{EXACT ALLOWED REGIONS}
\label{EXACT}

The construction
of exact confidence intervals
has been introduced by Neyman in 1937
(see, for example, Ref.\cite{Eadie-71,qun}).
It guarantees that the resulting confidence 
intervals have correct frequentist coverage, i.e. they belong to a 
set of confidence intervals obtained with different or similar, real
or hypothetical experiments that cover the true values of the parameters
with the desired probability given by the chosen confidence 
level
(see, for example, Ref.\cite{qudu}).
We apply this method in order to find confidence intervals with proper
coverage for the neutrino oscillation parameters.

Starting with the choice of an appropriate estimator of the parameter
under investigation,
for any possible value of the parameter one 
calculates an {\it acceptance interval} with probability $\beta$,
i.e. an interval 
of the estimator that contains $100\beta$\% of the values of the estimator 
obtained in a large series of trials.

Once the $100\beta$\% acceptance interval for each possible value of the 
parameter is calculated, the $100\beta$\% {\it confidence interval} 
is simply composed by all the parameter values whose acceptance interval 
covers the measured value of the estimator.

This procedure can be generalized to the case of more parameters:
the acceptance intervals (and also the confidence intervals)
are multidimensional regions and could be composed by disjoint
subintervals.

In the case of solar neutrino oscillations,
we have two parameters, $\Delta m^2$ and $\tan^2{\theta}$,
estimated through the minimum of $X^2$.
We consider a grid with 5000 points in the MSW region
and 
6000 points in the VO region.
For each point of the
grid we ge\-ne\-ra\-te about $6.5 \times 10^4$
synthetic data sets
that allow to calculate the distribution of the estimator
$X^2_{\mathrm{min}}$
and the consequent acceptance regions.
More details have been presented in Ref.\cite{noi-mc}.

\begin{figure}[t!]
\begin{center}
\rotatebox{90}{\includegraphics[bb=100 309 540 741,width=0.49\textwidth]{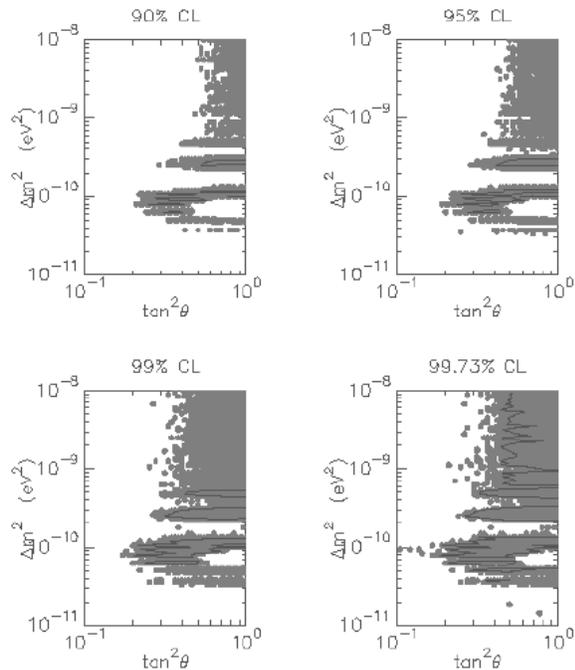}}
\end{center}
\caption{Allowed $90\%$, $95\%$, $99\%$, $99.73\%$ CL regions in the
VO part of the
$\tan^2{\theta}$--$\Delta m^2$ plane.
The gray areas are the allowed regions with exact frequentist co\-ve\-ra\-ge in
the VO region; the areas enclosed by the solid lines are the standard VO allowed
regions.}
\label{fig:vuocin}
\end{figure}

Our results are shown in Figures~\ref{fig:mswqua}
and
\ref{fig:vuocin},
where we have plotted the
$90\%$, $95\%$, $99\%$, $99.73\%$ CL regions in the
MSW and VO parts of the
$\tan^2{\theta}$--$\Delta m^2$ plane.
One can see that
the exact allowed regions are much larger than the standard ones.
In particular,
the LMA, LOW and VO regions are connected,
giving an allowed band at large mixing angles
for all $\Delta{m}^2 \gtrsim 10^{-10} \, \mathrm{eV}^2$.
Only the standard SMA region is
a good approximation of the exact one.

\section{CONCLUSIONS}

We have calculated with Monte Carlo the goodness of fit and the confidence level 
of the standard allowed regions of the two-neutrino oscillation parameters
$\Delta m^2$ and $\tan^2{\theta}$
obtained from the fit of the total rates
measured by solar neutrino experiments.

As expected,
we found that
the standard method overestimates the goodness of fit and
the confidence level
of the standard allowed regions.

Using Neyman's construction,
we have calculated exact allowed regions
with correct frequentist coverage.
We have shown that
the exact VO, LMA and LOW regions are much larger than the standard ones
and merge together
giving an allowed band at large mixing angles
for all $\Delta{m}^2 \gtrsim 10^{-10} \, \mathrm{eV}^2$.

\end{document}